\documentclass[letterpaper, 10 pt, conference]{ieeeconf}  

\IEEEoverridecommandlockouts                              

\usepackage{amsmath}
\usepackage{graphics}
\usepackage{multirow}
\usepackage{svg}
\usepackage{xcolor}
\usepackage{xspace}
\usepackage[pagebackref,breaklinks,colorlinks]{hyperref}
\usepackage{booktabs}
\newcommand{\sssec}[1]{ {{\flushleft \textbf{#1}}}}

\newcommand{\name}{Umbra\xspace}

\title{\LARGE \bf
Millimeter Wave Inverse Pinhole Imaging
}

\author{Akarsh Prabhakara, Yawen Liu, Aswin C. Sankaranarayanan, Anthony Rowe and Swarun Kumar
\thanks{A. Prabhakara is with University of Wisconsin - Madison, Madison WI 53706. Email: {\tt\small akarsh@cs.wisc.edu}. Y. Liu, A. C. Sankaranarayanan, A. Rowe, and S. Kumar are with Carnegie Mellon University, Pittsburgh PA 15213.  Email: {\tt\small \{yawenl@, saswin@, agr, swarunk\} @andrew.cmu.edu}}%
\thanks{This work has been submitted to the IEEE for possible publication. Copyright may be transferred without notice, after which this version may no longer be accessible.}
}

\begin{document}

\maketitle
\thispagestyle{empty}
\pagestyle{empty}


\begin{abstract}

Millimeter wave (mmWave) radars are popular for perception in vision-denied contexts due to their compact size. 
This paper explores emerging use-cases that involve static mount or momentarily-static compact radars, for example, a hovering drone. 
The key challenge with static compact radars is that their limited form-factor also limits their angular resolution. 
This paper presents Umbra, a mmWave high resolution imaging system, that introduces the concept of rotating mmWave “inverse pinholes" for angular resolution enhancement. 
We present the imaging system model, design, and evaluation of mmWave inverse pinholes. 
The inverse pinhole is attractive for its lightweight nature, which enables low-power rotation, upgrading static-mount radars. We also show how propellers in aerial vehicles act as natural inverse pinholes and can enjoy the benefits of high-resolution imaging even while they are momentarily static, e.g.,  hovering. Our evaluation shows Umbra resolving up to 2.5$^{\circ}$ with just a single antenna, a 5$\times$ improvement compared to 14$^{\circ}$ from a compact mmWave radar baseline.

\end{abstract}

\section{Introduction}
\label{sec:intro}

\begin{center}
\vspace*{0.1in}\textit{``The eye is always caught by light, $\qquad\qquad\qquad\qquad$\\ but shadows have more to say.'' --- Gregory Maguire}
\end{center}

Compact millimeter wave (mmWave) radars are popular in mobile setups like automotive and robotic perception under vision-impaired conditions \cite{pan2024ratrack,prabhakara2023high}, static wall-mount radars for seeing around corners \cite{dodds2024around}, pole-mount radars for wide area surveillance in farm lands, building premises, parking lots \cite{guan2020hawkeye}, and wall-mount in homes  \cite{hunt2024radcloud}. The challenge with compact mmWave radars is their poor angular resolution, compared to cameras and lidars with two to three orders of magnitude higher resolutions. This greatly restricts them to only coarse-grained sensing tasks, as opposed to full-fledged imaging. This fundamental limitation arises from the size of the imaging aperture. As determined by the Rayleigh criterion \cite{zhou2019modern}, the angular resolution ($\propto\lambda/D$, $\lambda$ is wavelength and $D$ is aperture) is poor due to compact radars having few ($<$ 10) antennas and small form factors (1-2 cm) \cite{awr}. 

Enhancing resolution has always involved expensive high-complexity antenna arrays with large apertures \cite{awrcascade} or physically moving a cheap compact radar to synthetically increase aperture \cite{yanik2019near,prabhakara2020osprey,qian2020} called Synthetic Aperture Radar (SAR). SAR is appealing for reducing cost and useful in cars and robots where radar motion is natural. In this paper, \textit{we are interested in scenarios that demand high angular resolution despite  restricted radar motion}: the aforementioned static wall or pole mount radar use cases or when a drone, with a radar, loses high quality imaging resolution while momentarily hovering in-place and inspecting a scene. To address these, we either need to mount expensive antenna arrays or perform SAR with a dedicated motion stage that continuously moves the radar. 
This raises the following question: Is it possible to avoid explicit radar motion, but still obtain the high angular resolution of SAR?

\begin{figure}
\centering
\includegraphics[width=\columnwidth]{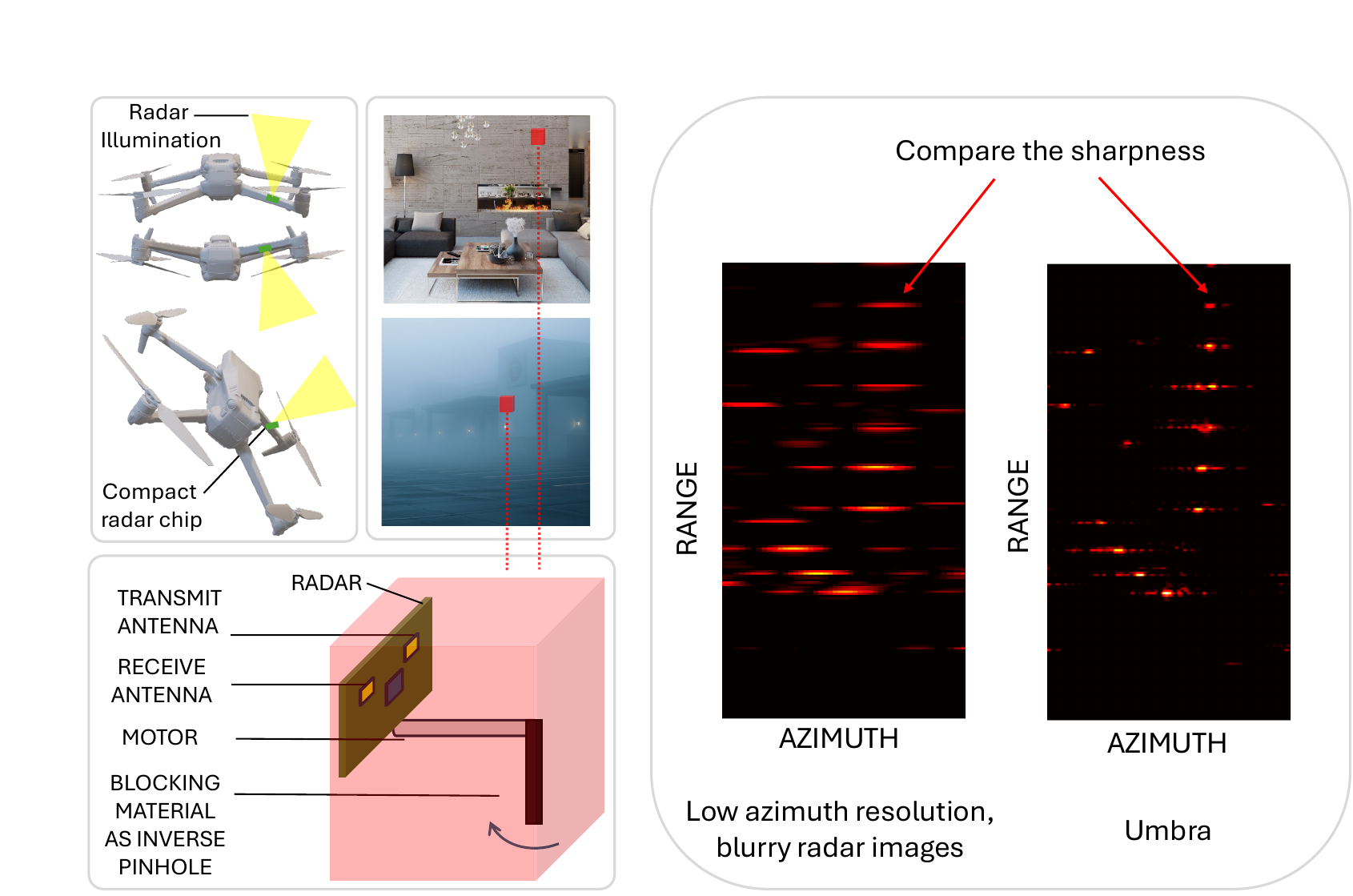}
\caption{\name\ designs an inverse pinhole to image from a single antenna on a mmWave radar for static mount radars. For momentarily static use cases like a drone hovering (puller configuration, pusher configuration, thrust vectored tilt and hover), we can exploit radar illuminating through rotating propellers as natural inverse pinholes. We show \name's superior angular (azimuth) resolution and sharpness compared to compact radars \cite{awr}.}
\vspace*{-0.25in}
\label{fig:hero}
\end{figure}

We propose a novel system that draws on a principle commonly explored in cameras --- the pinhole effect. To understand how it is related to resolution enhancement, let us move away from radars and consider a single photodiode. As a thought experiment, let us move a fully opaque sheet, with a tiny hole, in the field of view of the photo diode. Although a single photodiode does not have any angular information, the light that passes through the tiny hole is only from a specific angular portion of the scene. The movement of the tiny hole focuses on different angular portions over time, encoding spatial angular information temporally. Pinhole and more advanced masks are used in visible light cameras for depth estimation \cite{levin2007image} and in X-rays for imaging \cite{lu1995hard}. With radios traditionally operating on tens of centimeters and meters of wavelength, and pinhole sizes scaling with wavelengths, the masks are unreasonably large and bulky. The trend in commodity radios moving to higher frequencies (lower wavelengths) of mmWave (mm $\lambda$) presents a golden opportunity to create masks of dimensions that are practically reasonable. Thus, akin to a single pixel system, from just a single radar antenna we exploit this effect --- asking what mmWave pinholes look like and building the first 77 GHz (mmWave) pinhole imaging system.

We present \name\ (see Fig \ref{fig:hero}), a pinhole-based mmWave imaging system that uses the rotation of a lightweight object located very close to a \textit{single antenna}. This enhances angular resolution of static radars to 2.5$^{\circ}$, a 5$\times$ improvement over an off-the-shelf compact mmWave radar \cite{awr}. We first show how static mount radars can be upgraded with a custom-made rotating pinhole. Then, we show how propellers in aerial vehicles act as natural pinholes at mmWave frequencies. In building \name, we come across two key insights that make mmWave pinholes different from visible light or X-rays.

\sssec{Inverse pinhole design: } A pinhole typically has vast amounts of blocking material and a tiny transparent hole. The heavy blocking material is not suitable for moving fast. An inverse pinhole is the opposite of a pinhole, which lets all signals go through except a tiny blocker. When the blocker's shadow aligns with the object's angle (Fig \ref{fig:model-main}d), we see a dip in energy, as opposed to peak. The temporal occurrence of the dip encodes angle information. Because of less blocking material, inverse pinhole is an ideal, lightweight option, while still having the same resolution as pinhole. However, visible light sensors are limited by shot noise \cite{cossairt2012does}, where noise increases with signal amplitude. Inverse pinholes are flooded with a higher amplitude signal than regular pinholes. As a result, inverse pinhole, by and large, is not seen in visible light systems except perhaps in ``accidental'' settings \cite{torralba2014accidental}. mmWave not only presents a chance to build reasonably sized radio frequency pinholes but also presents a different sensor noise regime with signal independent thermal noise \cite{radionoise}. In this new regime, inverse pinhole is a feasible design and this makes the mask extremely lightweight to rotate with less motor power (Fig \ref{fig:hero}). 


\sssec{Bidirectional pass-through system model: } Unlike X-ray or vision, mmWave radars are active illumination imaging sensors where the transmit and receive antennas are closely located. The wave goes through the inverse pinhole twice inevitably --- starting from the transmitter, through the pinhole, backscattering from an object, through the pinhole again, to the receiver. We first show how this bidirectional pass-through is a positive phenomenon improving resolution. Second, our inverse pinhole being in the antenna near field and practical mask element dimensions being comparable to the wavelength, we take into consideration propagation models that account for diffraction effects as opposed to a simple shadow approximation that ray optics in a pinhole camera suggests. We then present robust reconstruction algorithms to recover the image under radar noise. 

We finally present material choices to realize effective inverse pinholes, imaging model accounting for non-idealities, and precise tracking of the motor at high update rates. With these, we demonstrate how static mount radars can be upgraded to higher resolutions. However, for momentarily static cases like a hovering drone, we cannot add custom material as it would damage a drone's aerodynamics. Instead, we show that the propellers (as is) are natural inverse pinholes, and enhance resolution.

\sssec{Contributions: }\name\ makes four key contributions.
\begin{itemize}
    \item A design based on mmWave inverse pinholes to enhance angular resolution with lightweight rotating loads.
    \item A system model dealing with the mmWave setup of bidirectional pass-through, diffraction effects, spherical wavefronts, antenna placement and radiation pattern. 
    \item mmWave inverse pinhole realization with material choice that just weighs 10 grams and the exploitation of drone propellers as natural inverse pinholes.
    \item A system evaluation on objects upto 20 meters away with 5x improvement over a compact radar baseline.
\end{itemize}

\section{Background}
\label{sec:background}

\begin{figure}
\centering
\vspace*{0.1in}
\includegraphics[width=0.9\columnwidth]{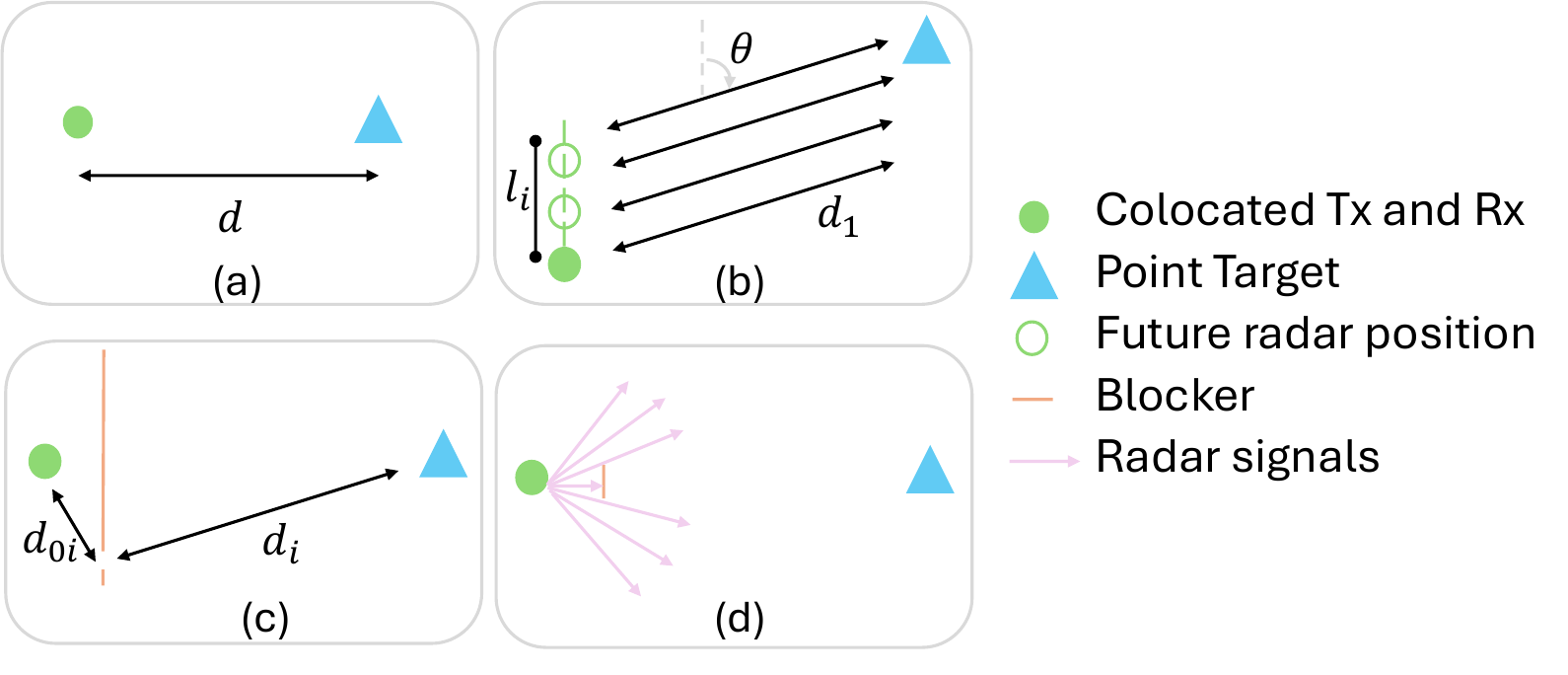}
\vspace*{-0.1in}
\caption{(a) shows a simple case of antenna and target; (b) shows a synthetic aperture radar case of a moving antenna; (c) shows a pinhole and (d) inverse pinhole imaging case with a static antenna}
\vspace*{-0.3in}
\label{fig:model-main}
\end{figure}

Radars use active illumination to obtain object-specific information like range, azimuth, elevation, and relative velocity. In the simplest setup (Fig \ref{fig:model-main}a), a colocated transmit (Tx) and receive (Rx) antenna transmits a wave, which reflects off a target and returns. The spherical outgoing wave decays as $1/d$, changes phase as $\frac{2\pi d}{\lambda}$; where $d$ is the one-way range; and $\lambda$ is mmWave wavelength (we use 77 GHz / 4 mm). One-way path from Tx to target is $\frac{1}{d}\exp{\frac{j 2 \pi d}{\lambda}}$ where $j=\sqrt-1$. The target spherically re-radiates back to the radar. Thus, measurements received at the radar are of the form $\frac{1}{d^2}\exp{\frac{j 2 \pi 2 d}{\lambda}}$ \cite{richards2010principles}. Imaging is essentially locating the target unambiguously in 3D (shown as 2D here). Clearly, a single antenna only has distance information, but no angles. 

Modern compact radars \cite{awr} have multiple non-colocated (8-12) antennas placed on a linear array to get angles. Another approach is to move and measure a single Tx/Rx antenna along a known trajectory (Fig \ref{fig:model-main}b). This is Synthetic Aperture Radar (SAR) processing \cite{richards2010principles}. With both approaches, we have $\frac{1}{d_i^2}\exp{\frac{j 2 \pi 2 d_i}{\lambda}}$, where $d_i$ is the instantaneous/antenna-specific distance. Essentially, the same target is seen from different (but known) antenna points of view. For a far target (plane wave assumed), the phase of these measurements is $\frac{2 \pi 2 (d_1+l_i \text{cos}\theta)}{\lambda}$, where $d_1$ is reference distance to the first antenna, $l_i$ is length from antenna 1 to $i$ and $\theta$ is the target azimuth angle for which we want to solve. With known antenna positions, we can solve for $\theta$. The classical bound on angular resolution is determined by the trajectory length $\lambda / (2*\text{trajectory length})$. Longer trajectories or array length (for a real antenna array) better the resolution. Our problem setup is focused on static deployment of popularly used cheap and compact radars \cite{awr_chip}. The compactness makes the default resolution poor. With static deployment, SAR requires powerful motors to spin radar. Thus, we look at other ways in which we can mimic SAR.

\section{Related Work}
\sssec{mmWave radar super-resolution: } Rotating an antenna can only provide a resolution of its field of view (FoV). For multi-antenna radars, resolution improves as $\propto\lambda/\text{Array Size}$. The TI single-chip radar \cite{awr} has 8 antennas (1.5 cm form factor) with 14.3$^{\circ}$ resolution. This results in coarse and blobby images. \cite{awrcascade} cascades 4 of these chips to obtain 86 antennas (11 cm form factor) and a resolution of 1.3$^{\circ}$. Other single-chip solutions achieve 2$^{\circ}$ resolution with 48 antennas \cite{vayyar}. Arrays of 10s of active antennas come at a significantly high cost. Software techniques to boost resolution on a cheaper, compact radar \cite{awr} are preferred. (1) Synthetic Aperture Radar: Radar motion along specific trajectories \cite{yanik2019near,prabhakara2020osprey,qian2020}; (2) Sensor Fusion with camera/lidar \cite{shuai2021millieye}; (3) Machine Learning / Sparse Processing Super-Resolution \cite{visualrecog24lai,guan2020hawkeye, prabhakara2023high}. In contrast, we seek to create dense images with just radar alone, without any training datasets or sparse target assumptions. Rather than moving a radar like in SAR, \name\ moves a lighter inverse pinhole.

\sssec{Pinhole and coded masks in other wavelengths: } Pinholes and masks in X-ray  \cite{lu1995hard} are used to estimate angles. In visible light they are used for depth imaging, high speed video, light field imaging, and lensless imaging \cite{levin2007image}. \cite{torralba2014accidental} explores ``accidental" inverse pinholes. For these sub-nm/nm wavelengths and pinholes of mm/cm dimensions, systems typically consider ray optics. Operating at mm-wavelengths, we consider diffraction and bidirectional pass. 

\sssec{Coded illumination in RF: } At Radio Frequency (RF), a large antenna array  is used to generate coded radiation patterns \cite{gollub2017large, yurduseven2017millimeter}. \cite{gopalsami2012passive,han2020millimeter} are unidirectional (transmit or receive only). \cite{sharma2021coded,sharma20223,smith2017security} are bidirectional. However, they vary in frequency (10-20, 35, 150 GHz) and use complex multi-hole codes needing active RF elements or moving metal sheets with holes. \cite{wang2025high} designs custom metasurfaces for each object being imaged. \cite{woodford2023metasight} builds metasurfaces away from the radar to extend coverage. We find that (1) bidirectionality is unique vs. optical and X-ray, (2) RF works \cite{sharma20223,sharma2021coded,smith2017security,wang2025high} lack an exposition on fundamental effects of bidirectionality, and (3) \name\ exploits simple, inverse pinholes that are \textit{both RF passive and lightweight}.


\section{\name System Design}

\name is motivated by a practical design point with a lightweight, inverse pinhole near a single transmit and receive antenna rotating at high speeds. This enables high resolution imaging 
with a low power motor. 
We choose rotation over linear actuation (that goes back and forth), to avoid losing momentum, minimizing acquisition time. This section first intuitively shows how we mimic SAR with pinhole motion, then presents an imaging system model accounting for mmWave bidirectional pass-through, then introduces the inverse design which makes the system lightweight, and finally discusses other practical system challenges including the natural inverse pinhole effect from propellers. 


\subsection{Pinhole motion to mimic SAR}
\label{ssec:design_space}

Our idea is to borrow and adapt the pinhole effect from vision or X-ray. Consider a pinhole sheet that is scanned in front of the radar (Fig \ref{fig:model-main}c), with an infinitesimally small pinhole (ignoring diffraction). This blocks all radar signals except where the pinhole opens. Essentially, this creates a virtual radar source at the pinhole location. Moving the pinhole sheet can be viewed as mimicking radar motion, creating different antenna points of view. At any scan position, for a known $d_{0i}$ dependent on the pinhole location and antenna, from $\approx\frac{1}{d_{0i}^2 d_i^2}\exp{\frac{j 2 \pi 2 (d_{0i}+d_i)}{\lambda}}$, we can compensate for $d_{0i}$. The rest of the terms look similar to the SAR setup seen in Sec \ref{sec:background}. Can we leverage this to do mmWave imaging with only a single antenna? Effective pinhole sizes increase on the scale of wavelength. At lower RF frequencies, 10s of cm or meter scale pinhole system would be impractical to scan. We leverage smaller mmWave wavelengths and move modest size (mm or cm scale) pinholes.

\subsection{mmWave Pinhole System Model}
\label{ssec:system}
This subsection builds an accurate imaging system model and simulator for mmWave pinholes to rapidly prototype pinhole designs and reconstruction algorithms. We want these unique attributes at mmWave to be captured: 
\begin{itemize}
    \item \textit{Bidirectional pass-through: } Because of closely separated Tx/Rx antennas (1 cm), a pinhole mask placed in the field of view of either of the antennas inevitably experiences both outgoing and incoming wavefronts.
    \item \textit{Wavelength-scale pinhole width: } A finite width pinhole of a modest size (a few mm) would be comparable to the wavelength. Rather than simply ray tracing, we need to account for diffraction effects in the shadow region. 
    \item \textit{Spherical wavefronts: } We envision locating the pinhole mask a few cm away. At this proximity, the wavefront is spherical when it impinges on the mask.
\end{itemize} 

To this end, we build a Rayleigh-Sommerfield \cite{goodman2005introduction} simulator at mmWave with all these requirements. In contrast to a one-way pass, the key difference is in the overall measurement model: $$y = (H\tilde{F} .* HF)x + n = Bx + n,$$
where $y$ is the measurements, $H$ encodes the time-variant blocking from a pinhole, $F$ and $\tilde{F}$ are the one-way propagation matrix for Tx to target and target to Rx, $x$ is reflectivity of targets in the scene that we want to recover, $n$ is additive noise, $.*$ is element wise multiplication. Further, the antenna has FoV and direction-dependent sensitivity determined by its radiation pattern. For each element in $F$, we multiply by a sensitivity term (from \cite{awr}). Given the proximity of the pinhole to the antennas and the sufficiently wide beamwidth of $\pm50 ^{\circ}$, the entire pinhole motion can be used effectively.

\sssec{Effect of bidirectional pass-through:} A unidirectional pass-through  only captures $H\tilde{F}x$. For bidirectional pass, let's simplify and assume colocated Tx and Rx ($\tilde{F}=F$). Let each element in $HF$ be of the form of $\approx\frac{1}{d}\exp{\frac{j 2 \pi d}{\lambda}}$, where $d$ is an equivalent distance term that captures one direction from antenna to target. Then, each element in bidirectional pass is $\approx\frac{1}{d^2}\exp{\frac{j 2 \pi 2 d}{\lambda}}$. To illustrate the significance of this, we draw an analogy from classical antenna array and SAR without any pinholes.

\begin{figure}
\centering
\vspace*{0.1in}
\includegraphics[width=0.9\columnwidth]{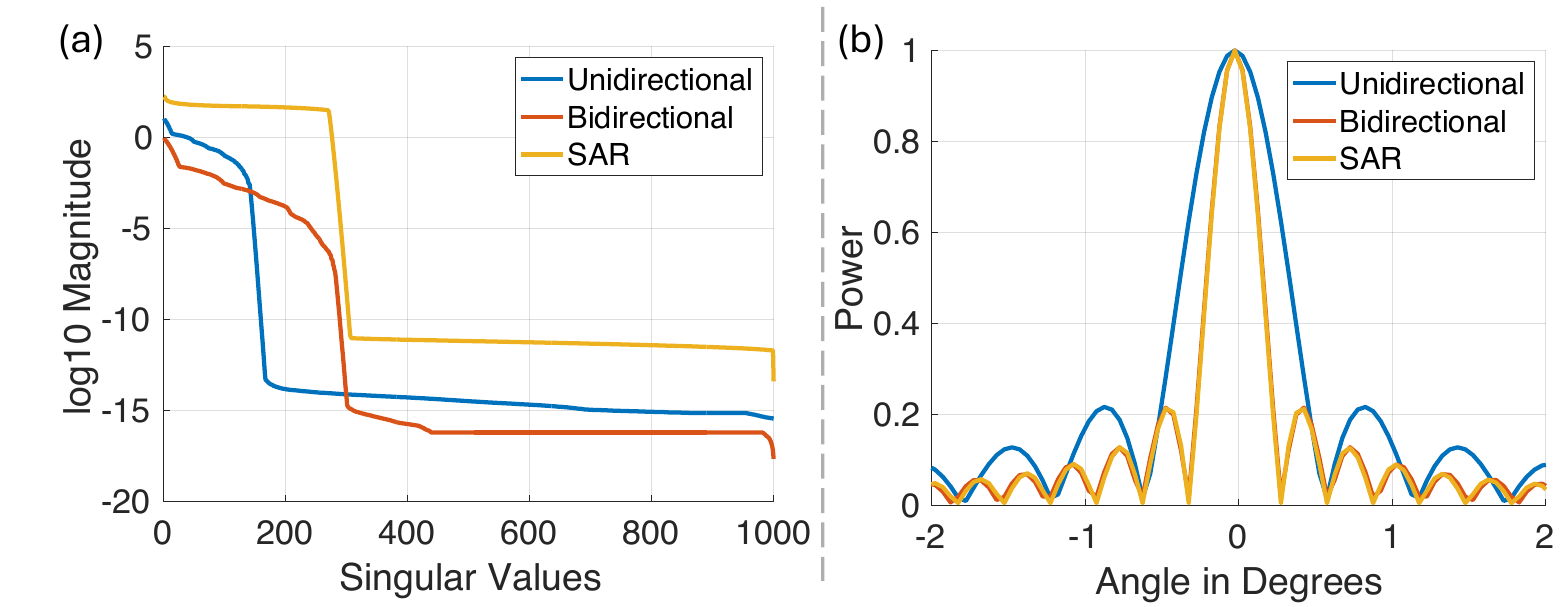}
\vspace*{-0.15in}
\caption{Shows the positive impact, on (a) singular values (b) angular resolution, of bidirectional pass in contrast to unidirectional pass.}
\vspace*{-0.3in}
\label{fig:coded_sar}
\end{figure}

Consider an antenna array with one static Tx and static linear receiver array. We have $\frac{1}{d_0 d_i}\exp{\frac{j 2 \pi (d_0+d_i)}{\lambda}}$, where $d_0$ is the constant distance from Tx to the target, $d_i$ is the distance from each array element to the target. Compensating for constants and looking at the relative phases between array elements, we end up with a form $\propto \frac{1}{d_i}\exp{\frac{j 2 \pi d_i}{\lambda}}$. Next, consider a SAR approach that moves both a colocated Tx and Rx resulting in $\frac{1}{d_i^2}\exp{\frac{j 2 \pi 2 d_i}{\lambda}}$, where $d_i$ is the instantaneous distance from Tx/Rx to target. Observing these two, we can draw equivalences between (1) unidirectional pinhole system $\equiv$ single Tx \& multiple Rx array (2) bidirectional pinhole system $\equiv$ single colocated Tx/Rx SAR. The $2d$ term is the main difference between these two forms and this implies that phase shifts are twice as sensitive with a bidirectional pass / SAR. In other words, SAR of length $L/2$ achieves identical phase shifts from a certain target as a receive-only array of length $L$, and thereby same resolution performance. Similar to this analogy, bidirectional pass-through should be more effective in enhancing the angular resolution than unidirectional. Thus, Tx and Rx sharing the pinhole mask turns out as a positive phenomenon.

We test out our analysis in simulation, and compare the singular values of the matrices $H\tilde{F}$ and $H\tilde{F} .* HF$ (Fig \ref{fig:coded_sar}a). The singular value curve stretching as far to the right at high magnitudes is ideal as it points to greater diversity in measurements and better resolution. We find the singular values drop off in significance much slower in the bidirectional case, thus more diversity. This leads to a $\approx$ 2x improvement in angular resolution (full width at half power) (Fig \ref{fig:coded_sar}b) matching our analogy explanation above. \\



\noindent\fbox{\begin{minipage}{0.97\columnwidth}
\textit{Take Away 1:} Bidirectional passthrough may seem harder to model vis-a-vis unidirectional passthrough but is actually beneficial and doubles angular resolution.
\end{minipage}}

\sssec{Comparison against SAR:} While the measurements are similar between bidirectional pinhole and SAR, we characterize our azimuth resolution  against circular SAR. In this, the antenna moves along the circumference of the same circle (dimensions same, since this dictates resolution), that the pinhole motion creates. Using all singular values, we achieve resolution on par with SAR (Fig \ref{fig:coded_sar}b). \\

\noindent\fbox{\begin{minipage}{0.97\columnwidth}
\textit{Take Away 2:} In the absence of noise, pinhole motion achieves resolution on par with SAR.
\end{minipage}}


 
\sssec{Robust reconstruction algorithm:} Ideally, simply inverting the matrix $B$ should reconstruct $x$. But in high noise, $B$'s pseudo inverse will amplify noise and reconstruction fails. Depending on the radar configurations, we may be in a high noise setting. To circumvent this, we resort to truncated SVD and then pseudo inverting $B$. Truncated SVD restricts the singular values to only dominant ones and reduces radar noise amplification. Restricting the maximum number of singular values ($\sigma_\text{MAX}$) adversely affects system resolution. The singular value decay pattern as seen in Fig \ref{fig:coded_sar}a is much sharper than the flat curve seen with SAR. Thus, while SAR gives good resolution even in high noise settings, restricting the number of singular values loses resolution. Fortunately, as seen in our evaluations, we are in a noise regime where several singular values (upto 40) can be used. This makes even simple pinhole designs very effective. \\

\noindent\fbox{\begin{minipage}{0.97\columnwidth}
\textit{Take Away 3:} Pinholes are unfortunately less resilient to noise vis-a-vis SAR, nevertheless offers significant resolution improvements to compact radars baselines.
\end{minipage}}

\begin{figure}
\centering
\vspace*{0.1in}
\includegraphics[width=\columnwidth,height=1in]{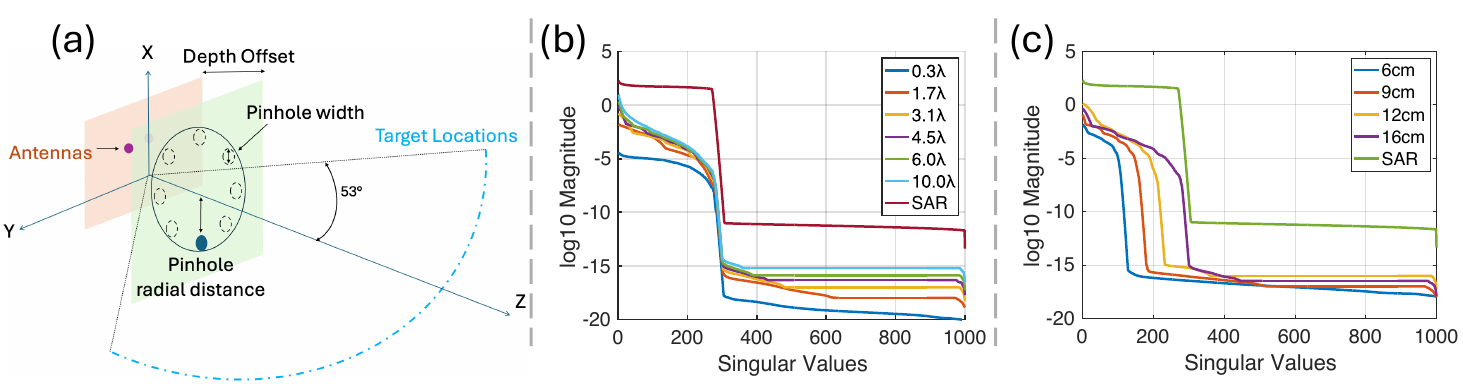}
\vspace*{-0.2in}
\caption{(a) shows pinhole setup. (b) \& (c) contrasts singular values by varying pinhole width and pinhole's location.}
\vspace*{-0.3in}
\label{fig:dims}
\end{figure}

\subsection{Pinhole Design}
\label{ssec:design_p}
We now know what a pinhole mask does to a single antenna to enhance resolution. This subsection focuses on pinhole design: geometry and the need for inverse pinholes. 

\sssec{Pinhole Geometry: } The simplest structure is literally a hole in the FoV. An infinite and zero-width pinhole either lets all or no waves through, lacking any angle-resolving capability. As we vary hole width (Fig \ref{fig:dims}b), we find that as size increases (1) a flat (SAR-like) curve becomes less flat (2) more energy passes through and singular values' magnitude increases. For noise resilience, we strike a balance between curve flatness decrease and magnitude increase, choosing pinhole width of 4$\lambda$ for our simulations and evaluations.

Pinhole center distance from the motor axis ($Z$) decides size of the circle that the motion creates. From bidirectional pinhole and SAR analogy, larger circle, better resolution. Thus, larger the pinhole center's distance, better the resolution. We verify in simulation (Fig \ref{fig:dims}c) and choose a radius of 16 cm for our evaluations with a SAR resolution of 0.35$^{\circ}$  balancing size practicality and resolution. 

The depth between pinhole plane and antenna plane together with the radiation pattern determine the effective field of view. That is, $tan^{-1}\frac{16 \text{cm}}{\text{Depth}}$. Outside this FoV, the pinhole's illumination will not directly fall, only weakly coupling the spatial location. For a modest FoV of $\pm53 ^{\circ}$, we use simulations to choose 12 cm as the offset. 


\sssec{Inverse Pinhole: } The pinhole is built from blocker material. Emulating a tiny $4\lambda$ hole needs a lot of material weight. Inverse pinhole is a concept that lets all waves go through except where the pinhole is. For this, we just need a blocker at the hole location. This reduces the weight drastically. It can be held to the rotating motor with a lightweight support. This can be as simple as a single-blade propeller made of mmWave-transparent material and blocker at the pinhole location. Remember, mmWave degrades little with thin, cheap, light materials like plastic. But first, why are they not as popular as regular pinholes? 

\begin{figure}
\centering
\vspace*{0.1in}
\includegraphics[width=0.95\columnwidth,height=1in]{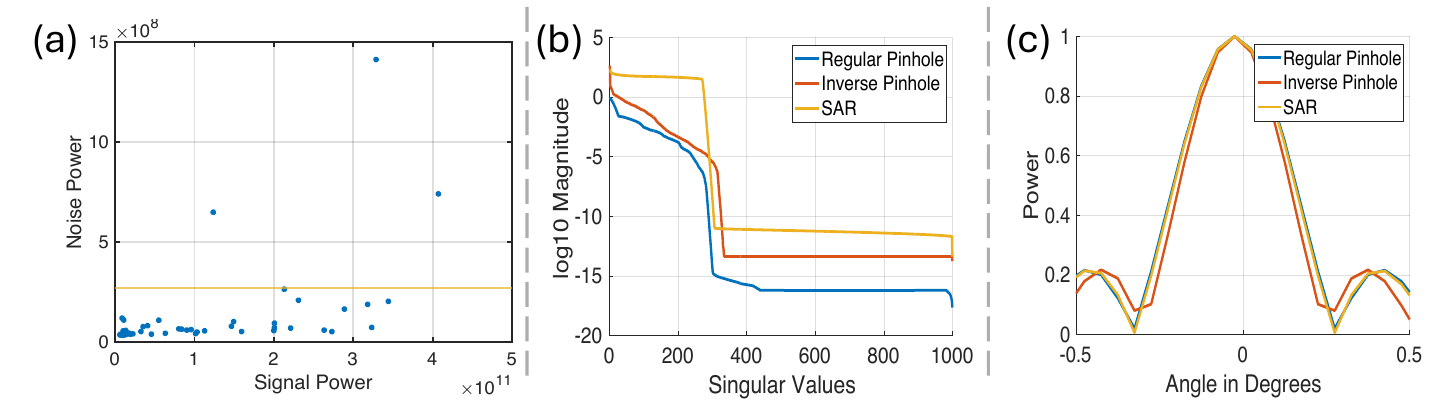}
\vspace*{-0.15in}
\caption{(a) Shows noise in mmWave radar isn't signal dependent. (b) \& (c) contrasts the singular values and angular resolution of regular pinhole and inverse pinhole respectively.}
\vspace*{-0.3in}
\label{fig:shot_noise}
\end{figure}

Visible light systems are dominated by a signal dependent shot noise \cite{cossairt2012does}. Inverse pinhole lets all light through except a small portion. This floods the visible light detector and enhances the noise dramatically. However, radio systems are dominated by signal independent thermal noise instead \cite{radionoise}. We verify the overall noise in our system from measurements. We experiment by moving a highly reflective target to different ranges from the radar. A closer target reflects more energy, creating higher amplitude signal at the receiver. However, irrespective of the target range (signal power), the noise power remains low and orders of magnitude lesser than signal power (Fig \ref{fig:shot_noise}a). The few outliers ($4\%$) that exist are due to signal dependent non-idealities of the radar that only show up when targets are extremely close and the signal power is high. This indicates that we are dominated by thermal noise. \\

\noindent\fbox{\begin{minipage}{0.97\columnwidth}
\textit{Take Away 4:} Inverse pinholes are attractive at mmWave due to signal independent noise regime.
\end{minipage}}

\sssec{Regular vs Inverse Pinhole: }Moving to inverse pinholes only causes 0s and 1s in $H$ to interchange. For a unidirectional pass, $HF$ and $(1-H)F$ are the models for regular and inverse pinholes. An inverse pinhole is equivalent to a regular pinhole upon background subtraction ($F$ is the background measurement without any pinhole). Thus, the resolution ability of both are similar. For a bidirectional pass (assuming colocated Tx/Rx), we have inverse pinhole as: $(1-H)F .* (1-H)F$, where, $OF.*OF$ is the background, $HF.*HF$ is regular pinhole both ways, $OF.*HF$ is regular pinhole for outgoing but open for incoming wave, $HF.*OF$ is vice versa, and $O$ is a matrix full of 1s. To understand the impact of these new terms, we use our simulator. We see that regular and inverse pinholes are not identical (Fig \ref{fig:shot_noise}b \& c) in a bidirectional setup. The singular value curve for the inverse pinhole lies strictly above the regular. Not only are inverse pinholes lighter, but given a noise setting, they can leverage a larger number of singular values above the noise magnitude and offer better resolution. \\




\noindent\fbox{\begin{minipage}{0.97\columnwidth}
\textit{Take Away 5:} For a bidirectional case, inverse pinholes offer better performance compared to regular pinholes. 
\end{minipage}}


\subsection{Practical Realization of Inverse Pinholes}

\begin{figure}
\centering
\vspace*{0.1in}
\includegraphics[width=0.8\columnwidth]{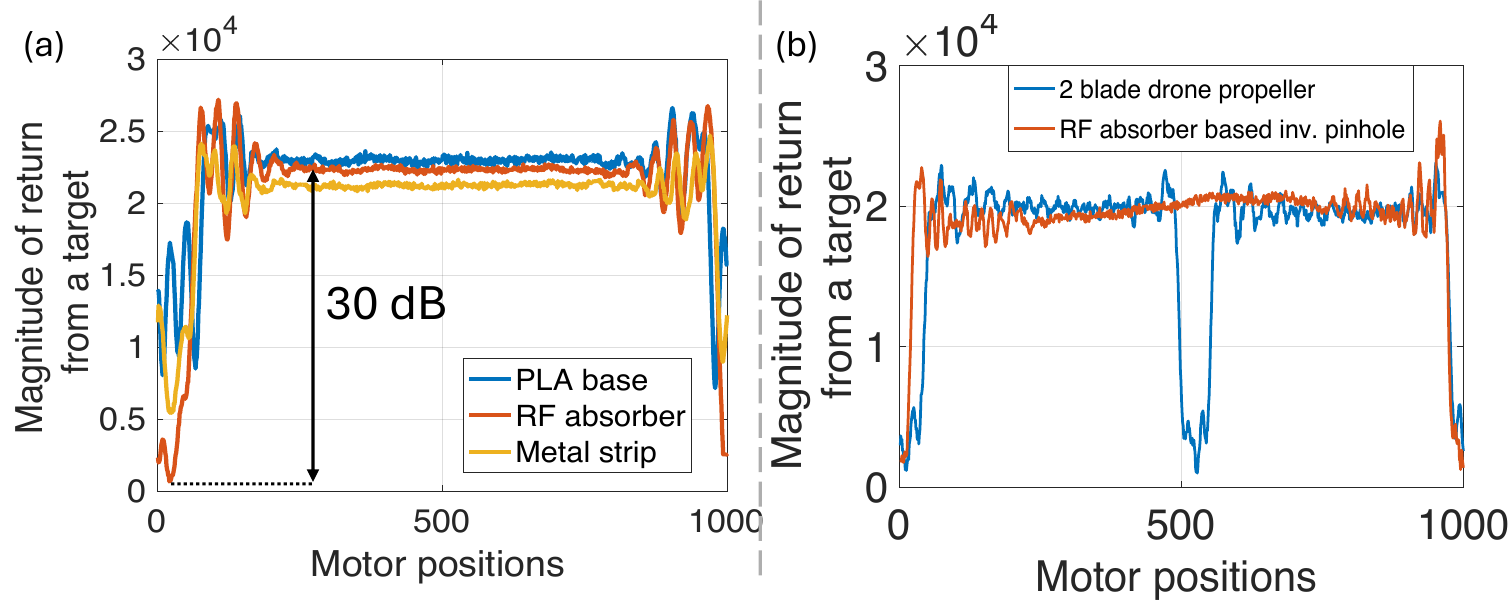}
\vspace*{-0.1in}
\caption{Shows return from a point target as an inverse pinhole completes 1 rotation (1000 positions). (a) Inverse pinhole material design choices. RF Absorber on a PLA base gives 30 dB attenuation (shadow). (b) Natural inverse pinhole effect from a 2-blade drone propeller yields two nulls.}
\vspace*{-0.3in}
\label{fig:system_ch}
\end{figure}

\label{ssec:practical}

\sssec{Blocker Material Selection: } To hold the inverse pinhole, we use a thin single-blade propeller-like supporting structure (16 cm length) of the same width (4$\lambda$) as the inverse pinhole. A natural design is to fill the entire support with blocker material (note that this is only along the radial direction, and is still much less material than emulating a pinhole). For effective inverse pinhole effect, we need effective blockers. A PLA plastic material causes a 9dB power attenuation (shadow) from a point target (see Fig \ref{fig:system_ch}a). A carbon fiber drone propeller's effects (with aerofoil geometry) is seen in Fig \ref{fig:system_ch}b. To a PLA substrate, we experiment with (1) absorbing material, (2) reflector (such as a metal strip) that is facing towards the radar. RF absorber creates the sharpest decrease in power (30dB) between the two. Metal strip only creates a 12dB power attenuation. We use the absorber on top of PLA for static-mount radar use cases and just the propeller's natural pinhole effect for the drone use case (Fig \ref{fig:system_ch}b). Due to airflow implications, we avoid adding blocker material to drone propellers and leverage the natural effect for this work. Future research can draw inspiration from this effect, design RF absorptive paint, balance extra weight and airflow. A calibration is performed with a point target to model matrix $B$ for each practical implementation's nuances: degree of blockage, aerofoil shape vs planar strip etc.


\sssec{Influence of Multi-blades: } The biggest difference between a drone's propeller and a custom-made inverse pinhole is the number of blades. Multi-blades create multiple nulls in one rotation (Fig \ref{fig:system_ch}b). The null location directly indicates target angle. But, if there are two nulls for a single target, this results in ambiguity (or aliasing). We can overcome this ambiguity by ensuring that the beam's spotlight passes through only one blade at any time. We can achieve this by building antennas with narrow elevation FoV (such as \cite{awr}) and offset from the center (like Fig \ref{fig:dims}a). Extending this to 4 or 6 blades involves shrinking the azimuth spotlight giving a reduced system FoV, a smaller ``useful" circle that is scanned, and thus poorer resolution. Using the elevation offset design, 2 blade propellers enjoy the natural inverse pinhole effect with same FoV and resolution as single blade.  


\sssec{High speed motor tracking: } Just like radar motion tracking in SAR, a pre-requisite for us is tightly synchronizing motor positions and radar measurements. Even drone motors constantly change rotation speeds. Motor encoders are a natural solution, but a high rate encoder to keep up with radar's 10 kHz drives up the motor's cost. Our solution is to use radar's high resolution (4 cm) ranging ability itself for self-synchronization. Placing the pinhole mask near the radar affects the first few range bins. The rotation creates a unique signature at these bins that periodically repeats and is scene-independent. If the motor rotates at non-uniform rates, this shows up in the signature. We continuously use Dynamic Time Warping on this signature to align multiple rotations. 

\sssec{Radar mounting on a drone: } Typically, radars look downward or sideways from a drone, suited for SAR. \name\ proposes to integrate mmWave antennas (with chip-scale packaging \cite{aop}) in the drone frame (akin to WiFi antennas on drones today) and orienting the beam to pass through the propeller (Fig \ref{fig:hero}). The beam orientation can be fixed (perpendicular to the plane of propeller) while the drone is tilted to the scene of interest, or it can be electronically oriented (with a 2 element transmit array). Note that \name's mounting and imaging is complementary and not a replacement to extremely long trajectory SAR as the drone moves. This is important as drones only hover momentarily. As the drone moves, the virtual pinhole positions scanned by a propeller move in a 3D space (as opposed to a planar circle). With drone pose estimation (used for SAR), we can jointly use measurements beyond in-place rotations and use a SAR + \name\ approach. This would ensure that we incur no downtime in high resolution imaging during hovering or moving. Here, we show \name\ with in-place hovering where SAR reliant on drone motion fails. The distance between the antenna and drone's propeller planes is kept at 12 cm, with the beam perpendicular to the planes. This frame-propeller distance is typical and doesn't significantly affect airflow.


\begin{figure}
\centering
\vspace*{0.1in}
\includegraphics[width=0.9\columnwidth]{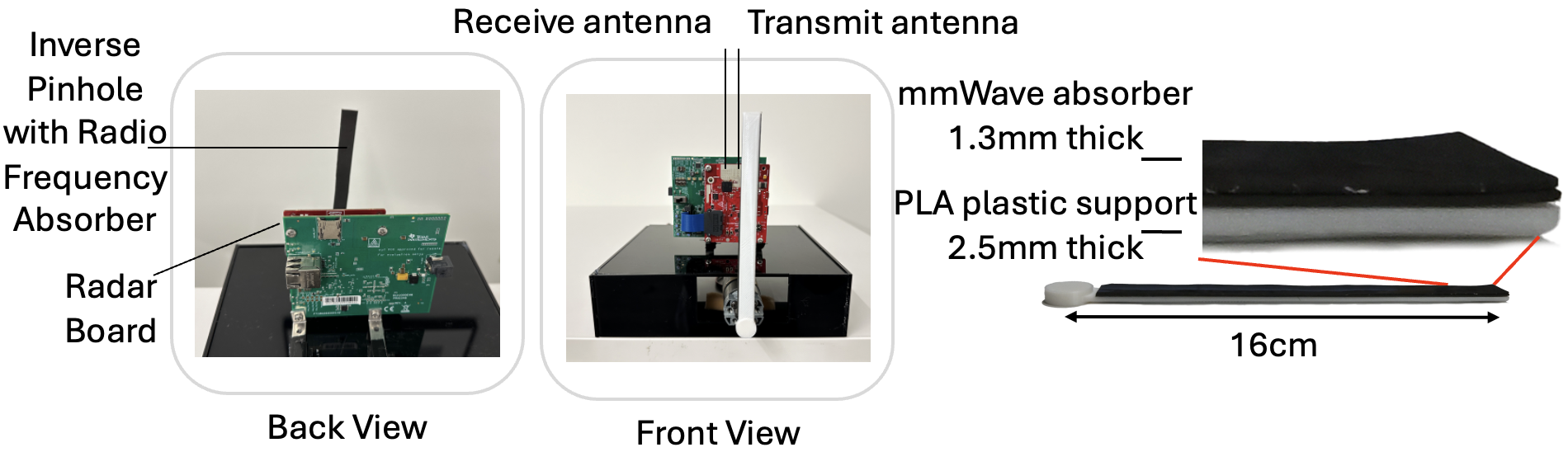}
 \vspace*{-0.1in}
\caption{Umbra moves a strip of radio absorbing material that just weighs 10 grams in front of a mmWave radar as seen in Fig \ref{fig:hero}}.
\vspace*{-0.3in}
\label{fig:impl}
\end{figure}

\section{Implementation}
\label{sec:impl}

\begin{figure*}[t]
\centering
\vspace*{0.1cm}
\includegraphics[width=0.97\textwidth]{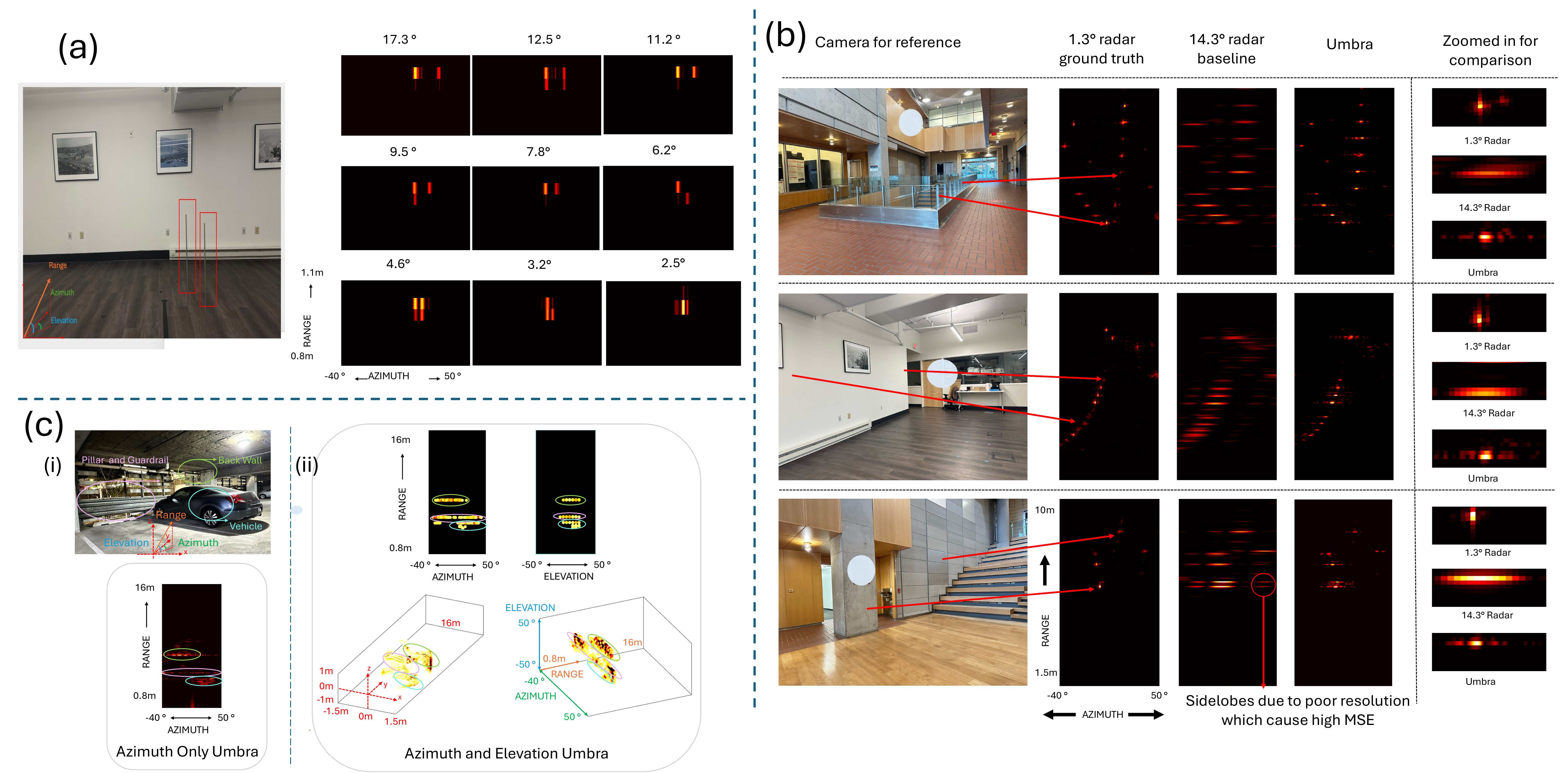}
\vspace*{-0.3cm}
\caption{(a) We characterize the azimuth resolution using two thin metal rods (in red boxes) as point targets. (b) Qualitative results show performance in static scenes. Intensity from [0.1,1] is mapped to colorscale. The rightmost column shows that we can generate sharper, high resolution images compared to our baseline. (c) Full 3D reconstruction of objects in a parking lot. Objects are marked with same color across different views.}
\vspace*{-0.4cm}
\label{fig:qual01}
\end{figure*}

\name\  uses AWR1843BOOST \cite{awr}, single-chip mmWave radar at 77-81 GHz that has 3 Tx and 4 Rx antennas. \textit{However, we just use a single Tx/Rx antenna pair}. This radar board is attached to DCA1000EVM for raw I/Q. We mount a cheap and low-end motor \cite{motor}. We place the inverse pinhole plane at a depth of 12 cm from the antenna plane, the antenna is 12 cm vertically offset from the rotation axis, and the motor rotates at 600 rpm. The inverse pinhole is a blocking structure that uses layers of RF-absorbing material \cite{absorber}, with a thin 3D printing PLA structure as a support (Fig \ref{fig:impl}). 
The total weight of our inverse pinhole is 10 grams while the radar board is 120 grams. To avoid exposing the spinning pinhole, for aesthetics or harsh environments, we can build a radome as shown in Fig \ref{fig:hero}. For showing natural pinhole effect, we use the DJI Matrice 100 drone and its default 2-blade propeller 16 cm long and minimum width of 1.5cm -- similar to (4$\lambda$) our custom designed inverse pinhole.


\section{Evaluation}

\begin{figure*}[t]
\centering
\vspace*{0.1in}
\includegraphics[width=0.97\textwidth]{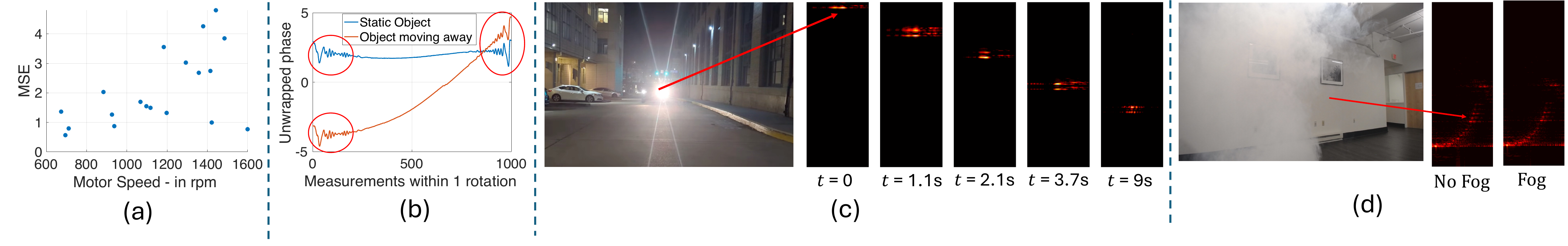}
\vspace*{-0.5cm}
\caption{(a) Error with motor speed. (b) Despite increasing phase shift due to object motion, the phase variations contributing to angle estimation (in red) are still encapsulated. (c) Qualitative result of a dynamic object like a car approaching a radar. The car's range decreases with time and the image quality is retained. (d) \name images show little error in seeing a wall with and without fog occlusion.}
\vspace*{-0.6cm}
\label{fig:qual2}
\end{figure*}

\sssec{Ground Truth and Baselines: }We build a rig with a co-located multi-chip and large form factor 1.3$^{\circ}$ azimuth resolution radar as ground truth \cite{awrcascade}. This is same as SAR over 8.5 cm trajectory length.  Our baseline uses all antennas on \cite{awr}. The co-location ensures overlap between GT radar and \name. All radars use Frequency Modulated Continuous Wave. Given wide ($\pm50^{\circ}$) FoV in azimuth, smaller elevation ($\pm20^{\circ}$) FoV and poor elevation resolution (18$^{\circ}$) in the GT, we mostly focus on range-azimuth (2D overhead view) imaging. Intensity indicates object reflectivities. We fix a range of 20 m. This can be changed for long range imaging.

\sssec{Metrics: } To compare resolution, we use a high frequency content estimator as a ``sharpness metric". Sharper the image, higher the energy in high frequency bands along the azimuth direction. For an image $I$ with $D(\text{range}_\text{freq},\text{azim}_\text{freq})$ as the 2D-FFT output, sharpness is $S(D(I)) = \sum|D(0, \text{azim}_\text{freq} >= 0.1)|$. The baseline's sharpness is finally defined as $\frac{S(D(\text{baseline image}))}{S(D(\text{ground truth}))}$. \name's sharpness is defined as $\frac{S(D(\text{\name's image}))}{S(D(\text{ground truth}))}$. Higher the value, closer is the sharpness to the ground truth.  For imaging accuracy, we use a pixel-wise mean squared error (MSE) on wide range of intensities in [0.01,1] that covers objects 40dB weaker in power. We also threshold images to convert to 2D point cloud and measure Chamfer Distance. Structural Similarity Index Measure (SSIM) shows perceptual quality.



\sssec{Resolution Tests: } For azimuth resolution tests, we use two point targets (6mm diameter metal rods). We move them on a precision stage from 17.3$^{\circ}$ until the two targets appear as one. Fig \ref{fig:qual01}a shows that we can resolve targets up to 2.5$^{\circ}$. For each measurement, the estimated angles align with the ground truth from the precision stage. Moving closer, at 1.5$^{\circ}$, we are unable to resolve the two targets. From our simulation, the resolution limit is 2.2$^{\circ}$ for our measured radar noise levels. This matches well with our measurements in reality.

\begin{table} 
\small
\centering
\vspace*{0.1in}
\begin{tabular}{@{}lcccc@{}}
\toprule
Method & Sharpness &  MSE & SSIM & CD \\
\midrule
\textbf{\name} & \textbf{0.80} & \textbf{13.11} & \textbf{0.85} & \textbf{0.19}\\
Compact radar baseline \cite{awr} & 0.64  & 21.20 & 0.82 & 0.25\\
\midrule
\textbf{\name} with $\sigma_\text{MAX} = 10$ & 0.70 & 18.69 & 0.83 & 0.20\\
\textbf{\name} with $\sigma_\text{MAX} = 15$  & 0.73 & 17.48 & 0.84 & 0.20\\
\textbf{\name} with $\sigma_\text{MAX} = 20$  & 0.73 & 18.61 & 0.82 & 0.21\\
\textbf{\name} with $\sigma_\text{MAX} = 30$  & 0.77 & 15.63 & 0.84 & 0.21\\
\textbf{\name} with $\sigma_\text{MAX} = 40$  & 0.80 & 13.11 & 0.85 & 0.19\\
\textbf{\name} with $\sigma_\text{MAX} = 60$  & 0.75 & 18.69 & 0.82 & 0.31\\
\textbf{\name} with $\sigma_\text{MAX} = 80$  & 0.60 & 30.12 & 0.72 & 0.37\\
\bottomrule
\end{tabular}
\vspace{-0.5em}
\caption{We evaluate on ``sharpness" (higher the better), MSE (lower the better), SSIM (closer to 1 is better), CD (in meters).}
\vspace*{-0.4in}
\label{tab:mean_ssim}
\end{table}

\sssec{Accuracy Tests in Natural Scenes: } We evaluate in office spaces, lobbies, and garages (open and cluttered spaces). Fig \ref{fig:qual01}b shows our superior imaging quality. It is easy to identify objects like metal posts and pillars. In the second scene in Fig \ref{fig:qual01}b, the wall appears discretized, as the studs behind the wall are strong reflectors. We used a stud finder to verify that the discretized points are indeed reflections from studs. We see the high quality reflected in all quantitative metrics, outperforming the baseline (see Table \ref{tab:mean_ssim}). Higher the $\sigma_\text{MAX}$ in Truncated SVD, better the resolution. This ablation trades-off resolution and noise. We find a sweet spot at $\sigma_\text{MAX}=40$. 


\sssec{Motor Dynamics: } We evaluate the impact of increasing motor speeds from the default 600 rpm for static objects. As we increase speed, for the same rate of radar data, the number of samples per rotation decreases, and the sensor body vibration increases. Vibration affects the phase of mmWave radars. Here, we empirically measure the MSE between the base image at 600 rpm and at higher speeds (Fig \ref{fig:qual2}a). Cross comparing from qualitative data, we find that the mild variance in MSE for points $<$ 1200 rpm, is due to lesser samples, and the high MSE for $>$1200 rpm is when the body vibration severely affects. Increasing radar's rate from 10 kHz, proofing motor vibration, and compensating for vibration in software will make \name\ robust. Our goal is to show the sensitivity for a certain implementation. 

\sssec{Object Dynamics: } We use a car driving into a parking lot at 5-10 mph as a moving test object. Note that other use cases in Sec \ref{sec:intro} involve human motion which is $<$3-5mph. Over one spin of the inverse pinhole, variance due to angle-dependent pinhole positions and object motion both influence measurements. Fig \ref{fig:qual2}b shows the increasing phase change as an object moves away added on top of the base phase variations due to object's angle. The same phenomena also affects SAR systems. However, if the object's doppler velocity is known, these effects can be easily compensated. The propagation model in Sec \ref{ssec:system} can be adapted to consider both object angle and velocity, to perform joint estimation. Focusing on a specific doppler velocity, and performing our algorithm, Fig \ref{fig:qual2}a shows that as the car approaches the radar, the object location approaches 0 m. The image quality holds good across frames. These images are generated at 10 fps with a 600 rpm motor. Higher object speeds demand dedicated motion compensation algorithms for inverse pinhole context which we leave for future work. Object phase shifts affects both SAR and \name, we just show we can achieve imaging by moving a lighter object. 



\sssec{Occlusion Tests: } Radar passes through occluders like fog with little effect (Fig \ref{fig:qual2}d). Mean difference in MSE with and without fog across images is 0.63.


\sssec{Full 3D Reconstruction: } Because we rotate as opposed to linear scan, we can obtain both azimuth and elevation resolution enhancements. In Fig \ref{fig:qual01}c(i), we first show range-azimuth images. Next, we modify the bidirectional propagation matrix to account for azimuth and elevation. Using this, we show range-azimuth slices, range-elevation slices, and range-azimuth-elevation images (Fig \ref{fig:qual01}c(ii)). We can see the car, back wall, and guard rails resolved in this 3D space. We resort to qualitative study as the ground truth radar has a poor 18$^{\circ}$ elevation resolution. Owing to specularity of metal objects, the images don't show the 3D shape well and are similar in quality to other azimuth-elevation SAR images \cite{guan2020hawkeye}.

\sssec{Multi-antenna vs SAR vs \name:} For 2.5$^{\circ}$ resolution, a multi-antenna array needs at least 45 antennas over 8.5 cm, with SAR (a single antenna) the radar board weighing 120g needs to spin 4.5 cm, with \name\ (also a single antenna) a 10g object is spun over 32 cm. SAR and \name\ reduce RF complexity, RF power and cost from that of 45 antennas. \name\ occupies a larger spin radius. However, power drawn is $mgr\omega$, $m$ is object mass, $g$ is gravity, $r$ is spin radius, $\omega$ is rotation speed. While \name's radius is 7x larger, SAR's mass is 12x larger, and thus, lower power with \name. While radar mass can be optimized to an extent \cite{aop}, for higher gain radars, radar mass dominates the power more. With \name, we present a new option to balance size-mass-power-cost. For systems constrained on \textit{RF complexity and power}, with \cite{awr} as the radar, \name\ offers low mass, power and cost, at a larger size (32 cm), affordable for wall and pole mounts. If a 4-8 element array \cite{awr} is available, \name\ (using 1 antenna) achieves the resolution of a 45 element array at the least. We can also use all available antennas (in $F$ and $\tilde{F}$), yielding higher SNR, usable singular values, and resolution.  

\begin{figure}
\centering
\vspace*{0.1in}
\includegraphics[width=\columnwidth]{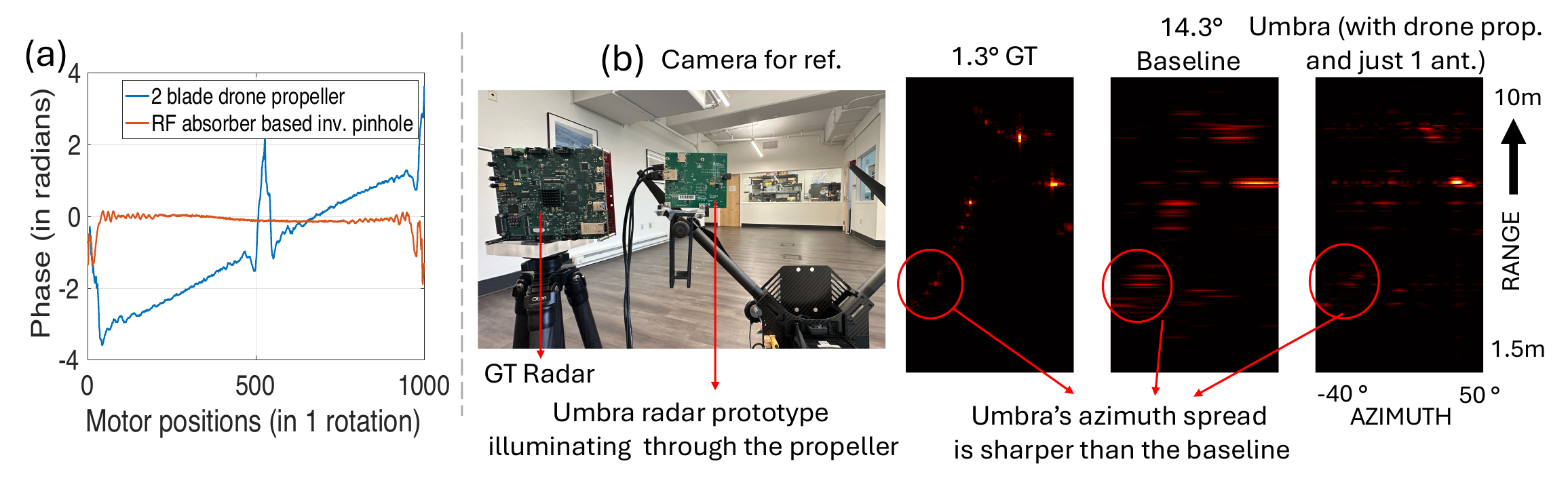}
 \vspace*{-0.3in}
\caption{Umbra exploits natural drone propeller rotation as inverse pinhole for resolution enhancement during hovering. The radar prototype used for Umbra can be integrated into the drone's frame as they can be chip-scale.}
\vspace*{-0.3in}
\label{fig:drone}
\end{figure}

\sssec{Natural Inverse Pinhole Effect:} Naturally rotating objects (e.g: drones, industrial fans etc.) can easily take advantage of \name. We show this with a static and tilted DJI Matrice 100 to mimic a hovering drone (in motor test mode \cite{dji}). A key challenge with drone propellers is a systematic phase change that arises due to the twin blade motion in addition to the inverse pinhole effect (Fig \ref{fig:drone}a shows the phase change for a point target). We estimate and cancel out this phase change before imaging. Fig \ref{fig:drone}b shows our performance using default propellers on the drone. The MSE for baseline and Umbra are 10.4 and 5.6 respectively. In practice, the radar would not stick out and would be integrated in the drone's frame. We choose a static evaluation instead of flying regular drones to point at interesting scenes (rather than the sky). We leave efforts to run \name\ on a flying drone (in pusher configuration or vectored thrust rotors) for the future.


\section{Conclusion}

\name\ paves the way to upgrade a single-antenna radar to perform high resolution imaging using pinhole motion for static or momentarily static use-cases. Unique to mmWave are bidirectional pinhole passes and the noise regime, which make inverse pinhole designs feasible. \name\ inspires creative exploitation of naturally rotating objects like drone propellers to serve as mmWave inverse pinholes.
\bibliographystyle{IEEEtran}
\bibliography{umbra_short}

\end{document}